# Ultrafast photo-magnetic recording in transparent medium

A. Stupakiewicz[1]*, K. Szerenos,[1] D. Afanasiev,[2] A. Kirilyuk,[2] A. V. Kimel[2]*

[1]*Laboratory of Magnetism, Faculty of Physics, University of Bialystok, 15-245 Bialystok, Poland.*
[2]*Institute for Molecules and Materials, Radboud University, 135, 6525 AJ, Nijmegen, The Netherlands.*

*Correspondence to: E-mail: and@uwb.edu.pl (A.S.); a.kimel@science.ru.nl (A.V.K.).*

**Finding a conceptually new way to control the magnetic state of media with the lowest possible production of heat and simultaneously at the fastest possible time-scale is a new challenge in fundamental magnetism[1-4] as well as an increasingly important issue in modern information technology[5]. Recent results demonstrate that exclusively in metals it is possible to switch magnetization between metastable states by femtosecond circularly polarized laser pulses[6-8]. However, despite the record breaking speed of the switching, the mechanisms in these materials are directly related to strong optical absorption and laser-induced heating close to the Curie temperature[9-12]. Here we report about ultrafast all-optical photo-magnetic recording in transparent dielectrics. In ferrimagnetic garnet film a single linearly polarized femtosecond laser pulse breaks the degeneracy between metastable magnetic states and promotes switching of spins between them. Changing the polarization of the laser pulse we deterministically steer the net magnetization in the garnet, write "0" and "1" magnetic bits at will. This mechanism operates at room temperature and allows ever fastest write-read magnetic recording event (<20 ps) accompanied by unprecedentedly low heat load (< 6 J/cm$^3$).**

In order to stabilize magnetic states of a single bit of a recording medium at room temperature, a magnetic anisotropy energy barrier of 60 $kT$ ~ 0.25 aJ (where $k$ is the Boltzman constant and $T$ is the absolute temperature) is taken as by far sufficient value[13]. This value would then also correspond to the energy which is ideally required to switch the magnetic state. In practice, however, about eight orders of magnitude more energy is used[14-16] and a lion share of it is subsequently lost via dissipation. It would be thus of great advantage to realize an optical switching of magnetic states, as light can be transferred with minimum losses and effectively modifies the barrier through opto-magnetic and photo-magnetic interactions[17,18].

Cobalt-substituted yttrium iron garnet (YIG:Co) is an optically transparent ferrimagnetic dielectric (see Methods and Extended Data Fig. 1) with cubic lattice and two antiferromagnetically coupled spin sublattices of $Fe^{3+}$ in both tetrahedral and octahedral sites[19]. The dopant $Co^{2+}$ and $Co^{3+}$ ions replace $Fe^{3+}$ in both types of sites[20]. These Co ions are responsible for a strong magnetocrystalline[21] and photo-induced magnetic anisotropy[18,22] as well as for a very large Gilbert damping[23] $\alpha$ = 0.2. In an unperturbed state at room temperature, the equilibrium orientation of the magnetization is defined by cubic ($K_1$ = –8.4×10$^3$ erg cm$^{-3}$) and uniaxial ($K_U$ = –2.5×10$^3$ erg cm$^{-3}$) anisotropy, which favor orientation of the magnetization along one of the body diagonals of the cubic cell (<111>–type of axes) and perpendicular to the [001] axis, respectively. It results in four easy magnetization axes which are slightly inclined from the body diagonals, as shown in Fig. 1. For easier distinction between different magnetic domains, we used a garnet film with a miscut of about 4º towards the [100] axis. In Figure 1, the large stripe-like domains have magnetizations along $M^{(L)}$+ near the [1-11] and $M^{(L)}$– near the [11-1] axes and the small labyrinth-like domains have magnetizations along $M^{(S)}$+ near the [111] and $M^{(S)}$– the [1-1-1] axes.

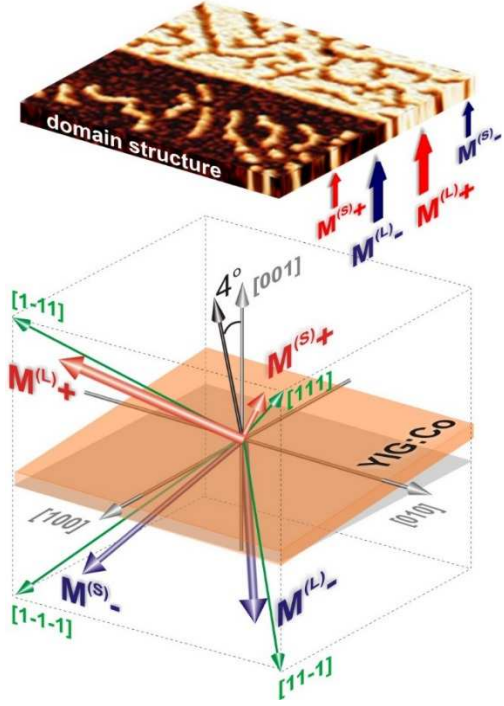

**Figure 1** | **Magnetic states and domain structure of YIG:Co.** Orientations of the easy magnetization axes and the pattern of magnetic domains with the magnetization directions close to [111], [1-11], [11-1] and [1-1-1] axes measured at zero magnetic field with magneto-optical polarizing microscope. Due to a miscut of the gadolinium gallium garnet substrate by 4°, the degeneracy between the states is partly broken and different magnetic domains cover non-equivalent volumes at zero external magnetic field[20]. The shown pattern with all four magnetization states ($M^{(L)}+$, $M^{(L)}-$, $M^{(S)}+$, $M^{(S)}-$) can be obtained as follows. First, the sample is brought into $M^{(L)}+$ state by an external magnetic field of $\mu_0 H = $ 80 mT applied along the [1-10] direction. Second, the field is removed and the sample turns into a state with $M^{(L)}+$ and $M^{(S)}-$ domains. Third, a magnetic field $\mu_0 H = 2$ mT for a short time applied along the [110] direction favors $M^{(L)}-$ domains and results in the final patter. After the magnetic field is removed, the pattern stays unchanged for at least several days. Magnetization orientations in the domains and type of the domain structure have been identified following the procedure explained in[20].

To investigate the feasibility of the switching with linearly polarized light in YIG:Co, we employed the technique of femtosecond magneto-optical imaging using pump pulse with duration of 50 fs (see Methods and Extended Data Fig. 2). The images of magnetic domains were taken before and after the excitation with a single pump laser pulse (see Fig. 2a). Taking the difference between the images underlines the photo-magnetic changes and is used for detailed analysis. Light can lift the degeneracy between the domains by generating photo-induced magnetic anisotropy[24]. In our case, pumping the initial pattern of magnetic domains with a single laser pulse polarized along the [100] axis ($\phi = 0°$) turns large white domains ($M^{(L)}+$) into large black ones ($M^{(L)}-$). Simultaneously, small black domains ($M^{(S)}-$) turn into small white ones ($M^{(S)}+$) (see Fig. 2a). The domain pattern stays remarkably unperturbed, only the contrast reverses. The initial state can be restored by pumping with a single laser pulse polarized along the [010] axis ($\phi = 90°$). The recorded domains are stable for several days due to the non-zero coercivity of the garnet film at room temperature (see Fig. 2b). The initial domain pattern can be also restored by a short-time application of an in-plane magnetic field of the order of 80 mT. The symmetry of the observed all-optical switching suggests that the point group of the crystal is *4* (see Supplementary). Although it is expected that the point group of the (001) garnet surface is *4mm*, the actual symmetry can be lowered either by the magnetization along the [001] axis or simply by distortions during the growth[25].

The minimum pump fluence required for the magnetic recording in YIG:Co is very sensitive to the wavelength of the pump pulse. The switched area estimated from the magneto-optical images is plotted as a function of the pump fluence for different pump wavelengths (see Fig. 3).

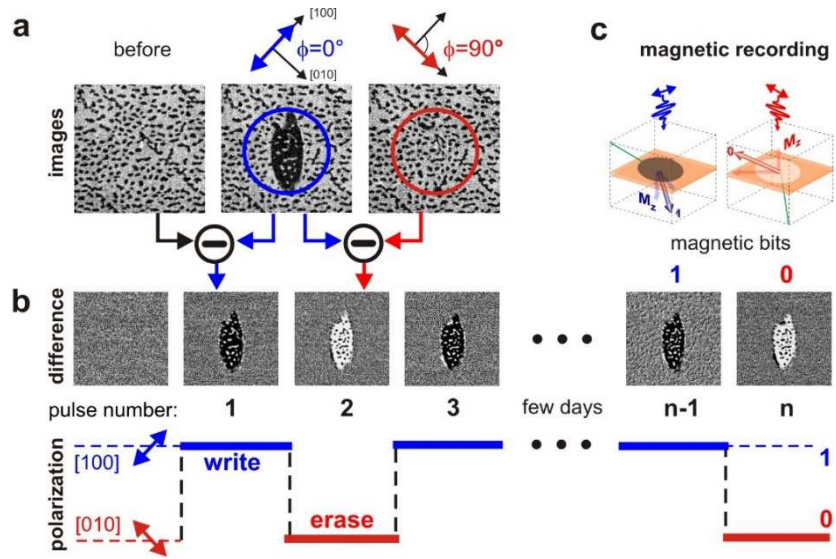

**Figure 2 | Single-pulse photo-magnetic recording.** The initial domain structure was prepared by applying an external magnetic field $\mu_0 H = 80$ mT along the [1-10] axis for few seconds. After removing this field a stable domain pattern was formed. The images are 200×200 µm² large. The pump beam with the wavelength of 1300 nm was focused to a spot of 130 µm in diameter and with maximum fluence of 150 mJ cm⁻². Panel (**a**) from left to right shows the domain pattern before the laser excitation, after excitation with a single laser pulse polarized along the [100] axis, and subsequent excitation with a similar laser pulse polarized along the [010] axis. Panel (**b**) shows differential changes after each of the pulse excitations and schematic demonstration of the ultrafast photo-magnetic recording of "0" and "1" bits with the linearly polarized pulses. Panel (**c**) schematically demonstrates the switching of the magnetization between two magnetic states $M^{(L)}+$ and $M^{(L)}-$ corresponding to all-optical recording of magnetic bits "0" and "1".

The wavelength was varied in the range between 1150 nm and 1450 nm (1.08–0.86 eV), where the light resonantly excites electronic transitions in Co-ions[26]. In the studied YIG:Co film, a resonant pumping of the $^5E \to {}^5T_2$ transition in Co-ions at the tetrahedral sites at 1305 nm ($\hbar\omega = 0.95$ eV)[27] is accompanied by absorption of about 12% ($a$=0.12) of light energy (Extended Data Fig. 1).

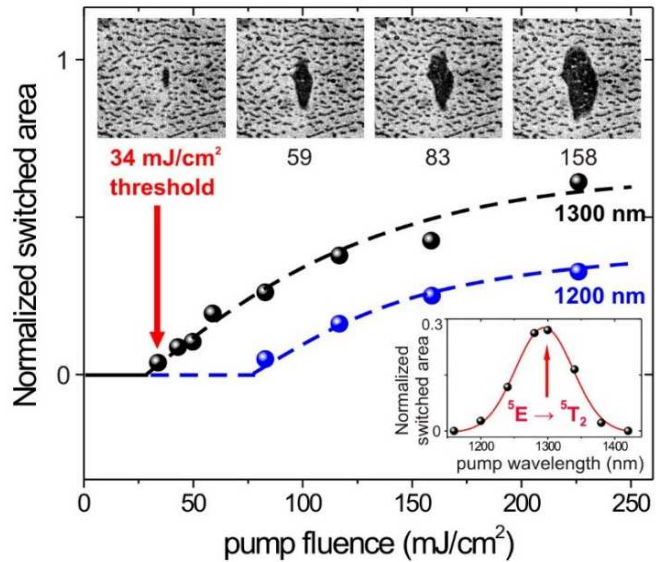

**Figure 3 | Energy efficiency of the all-optical magnetic recording.** The normalized switched area, calculated as the ratio of the recorded domain area (the black large domain on the images) to the area of pump laser spot $\pi r^2$ (where $r$ is the pump spot radius), is plotted as a function of the pump fluence. The plots correspond to the cases when the central wavelengths of the pump was around 1200 nm (blue dots) and 1300 nm (black dots and images). Dashed lines are guides to the eye. The magnetization is switched between $M^{(L)}+$ and $M^{(L)}-$ states (large domains) as well as between $M^{(S)}-$ and $M^{(S)}+$ states (small domains) with the help of the laser pulse polarized along the [100] axis (top panel of the images). The size of the images is 150×150 µm². The inset shows the spectral dependence of the normalized switched area for pump fluence of 83 mJ cm⁻³ (the red solid line was fitted by Gaussian function).

The spectral dependence in Fig. 3 reveals a pronounced resonant behavior around this transition. It is seen that the minimum pump fluence required to form a domain is about $I_{min}$=34 mJ cm$^{-2}$. It means that the magnetic recording is a result of absorption of about $aI_{min}/\hbar\omega \approx 3\times 10^{16}$ photons per cm$^2$. As the film is $d$ = 7.5 μm thick, the amount of absorbed photons required for the switching of the magnetization in a given volume would be about $10^{19}$ cm$^{-3}$ corresponding to depositing $aI_{min}/d \approx 6$ J cm$^{-3}$ of heat. For instance, recording a bit with the size 20×20×10 nm$^3$ would be accompanied by dissipations of just 22 aJ. To the best of our knowledge this is much lower than in the case of all-optical switching of metals 10 fJ[28] as well as in existing cases of hard disk drives 10–100 nJ[14], flash memory 10 nJ[15] or spin-transfer torque random–access memory 450 pJ–100 fJ[16].

Finally, we studied ultrafast dynamics of the magnetization switching employing time-resolved single-shot magneto-optical imaging (see Extended Data Fig. 2). The magnetic domains were recorded with a single pump pulse and imaged with a single 40 fs unfocused probe pulse with the central wavelength of 800 nm. After each write-read event the recorded domains were erased by application of an external magnetic field of 80 mT in [1-10] direction. Similarly to static magneto-optical imaging, reference images were taken before each pumping (i.e. at negative time delay). This image was subtracted from that obtained at a given pump-probe delay. Varying the time delay, series of the magneto-optical images were obtained (see inset top panel in Fig. 4). To quantify the dynamics of the laser-induced changes we took an integral over the pumped area, normalized the data and plotted the result as a function of time-delay between the pump and probe pulses. It is seen that the recorded domain emerges with the characteristic time $\tau$ of about 20 ps and gets stabilized after about 60 ps. As for the recording and reading out we used just two femtosecond laser pulses, to the best of our knowledge this experiment is the fastest ever write-read magnetic recording event[9]. Unlike all-optical magnetic switching in metals[6-11], the recording in transparent dielectrics does not require any destruction of magnetic order and operates without ultrafast heating of the medium up to the Curie point.

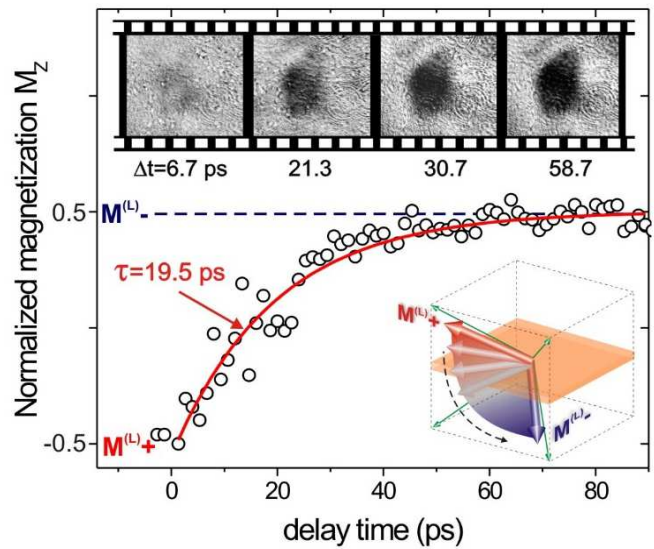

**Figure 4 | Time-resolved all-optical magnetic switching as observed by femtosecond single-shot imaging.** Time-dependence of the normalized magnetization projection on the [001] axis $M_Z$ with respect to the saturation magnetization. The data points were calculated as the ratio of the magneto-optical signal (the average image contrast) in the switched area to the magneto-optical signal in the case when the magnetization is aligned along the [001] axis. The red solid line was fitted using the exponential increase $(1-exp[-\Delta t/\tau])$ with the characteristic time $\tau$=19.5±1.6 ps. The top panel show the images of domains obtained at different time-delays after subtraction of the reference image obtained at negative $\Delta t$. The inset shows the schematics of the magnetization trajectory during the switching. The magnetization is switched between $M^{(L)}+$ and $M^{(L)}-$ states with the help of the laser pulse polarized along the [100] axis. The pump fluence was 150 mJ/cm$^2$ and the central wavelength of the pump was 1250 nm. The size of the images is 240×210 μm$^2$.

This time of the switching is in a very good agreement with the quarter of period of the laser-induced precession of magnetization in YIG:Co film (see Supplementary Fig. S1). Thus it is reasonable to suggest that, unlike all-optical switching in metals[9], the mechanism of the spin switching in garnets proceeds via the precession of the net magnetization. In this scenario in order to switch the magnetization from the initial $M^{(L)}+$ /$M^{(S)}-$ state, optical excitation should induce magnetic anisotropy which favors $M^{(L)}-$ and $M^{(S)}+$ states, respectively. It means that the magnetization will start precession around a direction somewhere in between these $M^{(L)}-$ and $M^{(S)}+$ states. When started from $M^{(L)}+$, after about 60 ps, i.e. after the first quarter of the precession period equal to 250 ps (see Supplementary Fig. S1a), the magnetization vector will be closer to $M^{(L)}-$ state. If at this moment due to relaxation of the photo-excited electrons the initial magnetic anisotropy is restored, the magnetization will start to process around the $M^{(L)}-$ direction eventually arriving to this metastable state on a time-scale defined by the damping of the oscillations. In reality in YIG:Co film, the life-time of the photo-induced anisotropy at room temperature is also of the order of 60 ps and the damping is indeed very large[23]. As a result, the magnetization in large domains moves along a trajectory from the initial $M^{(L)}+$ to the new metastable state $M^{(L)}-$ (see inset in Fig. 4). The switching of small domains between $M^{(S)}+$ and $M^{(S)}-$ occurs simultaneously and in the same fashion. From the orientation of the linear polarization of the pump light which results in the recording it can be concluded that the photo-induced anisotropy originates from optical excitation of Co ions at tetrahedral sites[22]. Note that the amplitude of the spin precession induced by light (see Supplementary Fig. S1b) is also at maximum when the linear polarization of light is along the [100] or the [010] axes thus supporting the above described mechanism.

We anticipate that the magnetization switching due to the photo-magnetic phenomena open up a plethora of opportunities for design and development of novel materials and methods for opto-magnetic recording. For instance, using the photo-magnetic garnet as a recording medium can anticipate a technology similar to Heat Assisted Magnetic Recording, but without a substantial heat and the need for an electromagnet. Furthermore, it is known that magnetic anisotropy in garnet films can be also controlled by electric fields[29]. Tuning the strengths of the magnetocrystaline, photo-magnetic and electrically induced anisotropies such that the switching is only possible under simultaneous action of the electric field and light seems to be a solution for the ever fastest and least dissipative Magnetic Random Access Memory.

**Acknowledgements** We acknowledge support from the National Science Centre Poland (grant DEC-2013/09/B/ST3/02669), the European Research Council under the European Union's Seventh Framework Program (FP7/2007-2013)/ERC Grant Agreement No. 257280 (Femtomagnetism) and the Foundation for Fundamental Research on Matter (FOM). We thank A. Chizhik, A. M. Kalashnikova for fruitful discussions as well as A. Maziewski and Th. Rasing for continuous support.

## METHODS

**Materials.** The magnetization switching results were obtained on $d$=7.5 μm thick Co-substitution yttrium iron garnet film (YIG:Co) with composition $Y_2CaFe_{3.9}Co_{0.1}GeO_{12}$. The single crystal YIG:Co garnet film was grown by liquid phase epitaxy on gadolinium gallium garnet $Gd_3Ga_5O_{12}$ (001)-oriented substrates with 4° miscut and thickness 400 μm. The saturation magnetization at room temperature was $4\pi M_S$=90 G and the Néel temperature was 445 K. The Gilbert damping measured using FMR technique gives $\alpha$=0.2. At room temperature the sample has both cubic ($K_I$= −8.4×10³ erg cm⁻³) and uniaxial ($K_U$= −2.5×10³ erg cm⁻³) anisotropy, which were measured by means of both ferromagnetic resonance and torque magnetometry. The cubic anisotropy term is dominating, yielding easy axes of magnetization along <111>-type directions. The uniaxial term modifies this by tilting the easy axes slightly towards sample plane. Thus such a crystal has eight possible magnetization states in zero applied field, directed close to body diagonals of a cube. By symmetry, those magnetization states should be energetically equivalent, but due to substrate miscut this degeneracy is lifted and some of them have slightly lower energy. This is why in the demagnetized state the sample shows an alternating stripe pattern of magnetic phases[20]. The measured absorption coefficient within spectral range of 1150-1450 nm for the studied sample is shown in Extended Data Fig. 1.

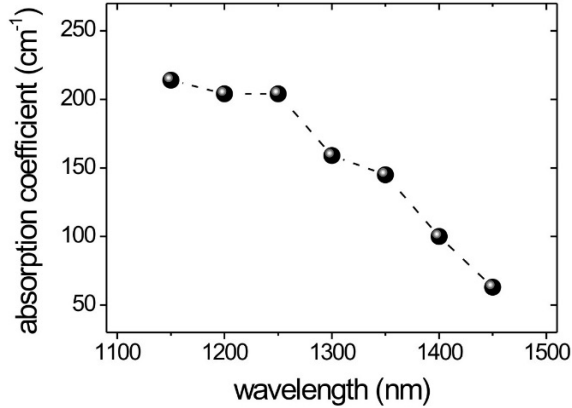

**Extended Data Figure 1 | Spectral dependence of absorption coefficient of YIG:Co film.**

**Experimental tools.** A design of the time-resolved pump-probe experimental set-up for investigations of all-optical magnetic recording is shown in Extended Data Fig. 2.

*A. Magneto-optical imaging under single pump pulse excitation.* The domain structure of the studied garnet films was visualized using magneto-optical polarizing microscope with standard LED source of polarized light as a probe. In this case, the central wavelength of pump pulse with duration of 50 fs was varied in the spectral range of 1150-1450 nm. Relying on the fact that domains with different orientation of the magnetization will result in different angle of the Faraday rotation, the domains were visualized with the help of an analyzer and a CCD camera. The images were acquired about 10 ms after excitation with a single pump pulse.

*B. Time-resolved femtosecond single-shot imaging.* For investigation of magnetization switching dynamics we employed laser pulses with duration of 40 fs and with the central wavelength of 800 nm as a probe. The linearly polarized unfocused probe beam passing through the sample was collected by an objective and the magneto-optical contrast of the sample was gained with a help of an analyzer (see Extended Data Fig. 2). The acquired magneto-optical image was digitized and recorded with the help of the CCD camera. A single shot pumping was achieved by placing a mechanical shutter in the path of the pump beam. The actuation time of the shutter was set to the minimum possible value of 60 ms. In order to exclude any possibility of excitation by more than one pump pulse, the repetition rate of the amplifier was brought down to 10 Hz. To improve signal-to-noise ratio in the detection of the probe pulses, the exposition time of the camera was set down to 1 ms. The activation time of the camera and the mechanical shutter were controlled by an electrical delay generator synchronized with the laser. A proper adjustment of the electrical delays for the shutter and the camera allowed to capture the magneto-optical image produced by a single pump pulse. The delay generator was set to the standby mode and was controlled by an external computer. The asynchronous trigger from the computer results in a generation of a single trigger signal synchronized with the laser which activates the camera and the shutter. To erase a long living state with the switched magnetization and to reinitialize the magnetic state, a magnetic field was applied after each single-shot event. Repeating such a single pump and single probe measurement for various values of the delays between the pump and the probe pulses (see Extended Data Fig. 2), we acquired images of the domains at various time moments after arrival of the pump pulse. The delay time $\Delta t$ could be adjusted in the range from 50 fs to 1 ns.

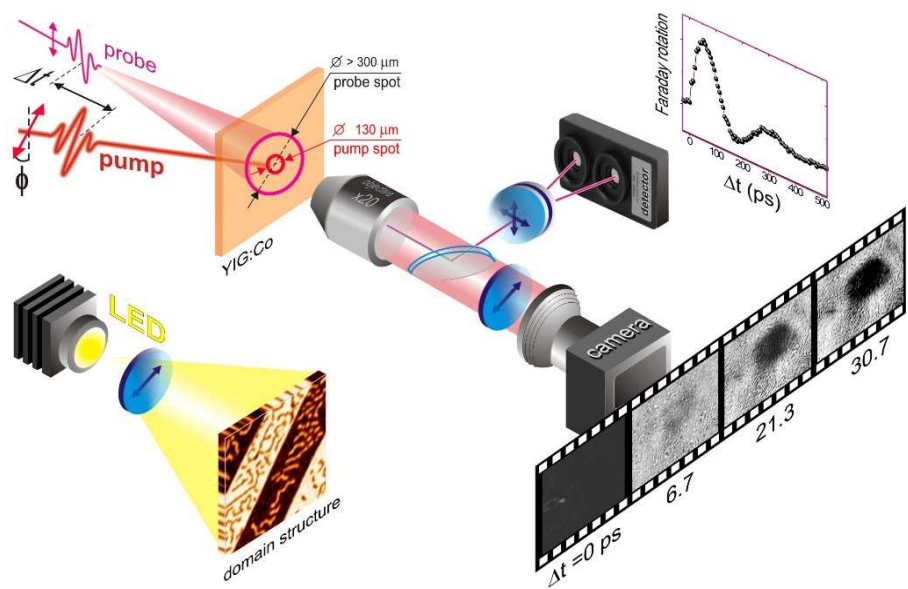

**Extended Data Figure 2 | Schematics of the time-resolved magnetization dynamics and single shot imaging.** Inset shows the magneto-optical visualization of the magnetic domains formed by a single laser pulse excitation of YIG:Co.

# SUPPLEMENTARY INFORMATION

**Estimation of temperature increase under laser pump pulse.** To estimate a temperature increase as a result of excitation by a single pump pulse in our experiment, we take into account the minimum intensity required for the switching $I_{min}$=34 mJ cm$^{-2}$, the heat capacity of the garnet $C$=430 J mole$^{-1}$ K$^{-1}$ at room temperature[19], the molar weight $m$=706 g mole$^{-1}$ and the density $\rho$=7.12 g cm$^{-3}$. The measured absorption of the pump at the wavelength of 1300 nm is about 12% ($a$=0.12, see Extended Data Fig. 1). The temperature increase as a result of absorption is thus $\Delta T = a \cdot I_{min} \cdot m / (C \cdot d \cdot \rho)$= 1.25 K. Such a temperature increase is at least two orders of magnitude lower than the one required to reach the Curie temperature.

**Laser-induced magnetization precession.** To study laser-induced spin oscillations induced by femtosecond laser pulses in YIG:Co film, we also carried out conventional time-resolved measurements using a magneto-optical pump-probe method. Pump pulses with duration of 50 fs arrived at the sample with the repetition rate of 500 Hz. The angle of incidence was set to 10° from the sample normal i.e from the [001] crystallographic axis of the garnet film. Equally short probe pulses had twice higher repetition rate and arrived at normal incidence to the sample. The central wavelength of pump beam was set to 1200 nm. This wavelength corresponds to the maximum amplitude of the spin oscillations. The central wavelength of the probe was 800 nm. The pump beam with fluence below 70 mJ cm$^{-2}$ was focused to a spot about 130 μm in diameter. The delay time $\Delta t$ between the pump and the probe pulses could be adjusted in the range from 50 fs to 1 ns. The polarization plane of the linearly polarized pump pulse was set at an angle $\phi$ with respect to the [100] axis. The polarization plane of the probe beam was along the [1-10] axis. Using balanced photodetector we measured the Faraday rotation of the probe as a function of the delay time $\Delta t$ between the pump and probe pulses (see in Extended Data Fig. 2). The Faraday rotation is proportional to the out-of-plane component of the magnetization $M_Z$. All measurements were done in zero applied magnetic field and at room temperature. The measurements were performed in a stroboscopic mode and thus revealed the pump-dynamics which is reproducible from pulse to pulse.

The magnetization precession signals show a strong dependence on the pump polarization (see Fig. S1a). Moving the pump spot across the boundary between large domains with $M^{(L)}+$ and $M^{(L)}-$ also allows to excite spin precession with opposite phases. It is also interesting to note that if probing spot is placed exactly on the wall, the antiphase signals from each domain average to zero. This is only possible if the underlying dynamics is due a coherent rotation of the magnetization and not due to domain wall motion. The characteristic rise time of the signal is about 20 ps. The time delay at which the signal is at maximum is about 60 ps (see Fig. S1a). Orthogonal polarizations along <100>-type directions result in the highest amplitude of the spin precession (see Fig. S1b). These polarizations correspond to the global symmetry axes of the tetrahedral in our garnet film hosting the Co ions. The amplitude of the precession follows a $sin(2\phi)$ dependence, which is typical for photomagnetic effects[18,22,24]. The amplitude of the magnetization precession increases linearly with the pump fluence due to stronger light-induced effective field (see Fig. S1b). The period of the precession is nearly constant (about 250 ps) and follows the frequency of the ferromagnetic resonance mode[23] in the field of magnetic anisotropy. Note that the central wavelength of the pump corresponding to the most efficient switching does not coincide with the wavelength corresponding the most effective excitation of the oscillations. This is due to the fact that the oscillations are detected in a stroboscopic mode, which gives the largest signal when no switching takes place. The microscopic mechanism of the photoinduced anisotropy has been extensively discussed before[18,22,24]. The charge transfer process redistributes electrons among Co ions in nonequivalent crystal sites, changing the valence states of the Co ions and consequently, their contribution to magnetic anisotropy. The latter can be explained in terms of single ion anisotropy model[21]. The pump light is used to excite optical transitions in the YIG:Co film in the tetrahedral Co ions. This approach allows to achieve a significant amplitude of the effective field of the photo-induced anisotropy (hundreds of Oe). Another important feature is that even though the number of the excited

Co ions does not reach 100%, the single-ion anisotropy from Co is very high, producing a large light-induced field. Thus the photomagnetic effect in the YIG:Co film is very efficient.

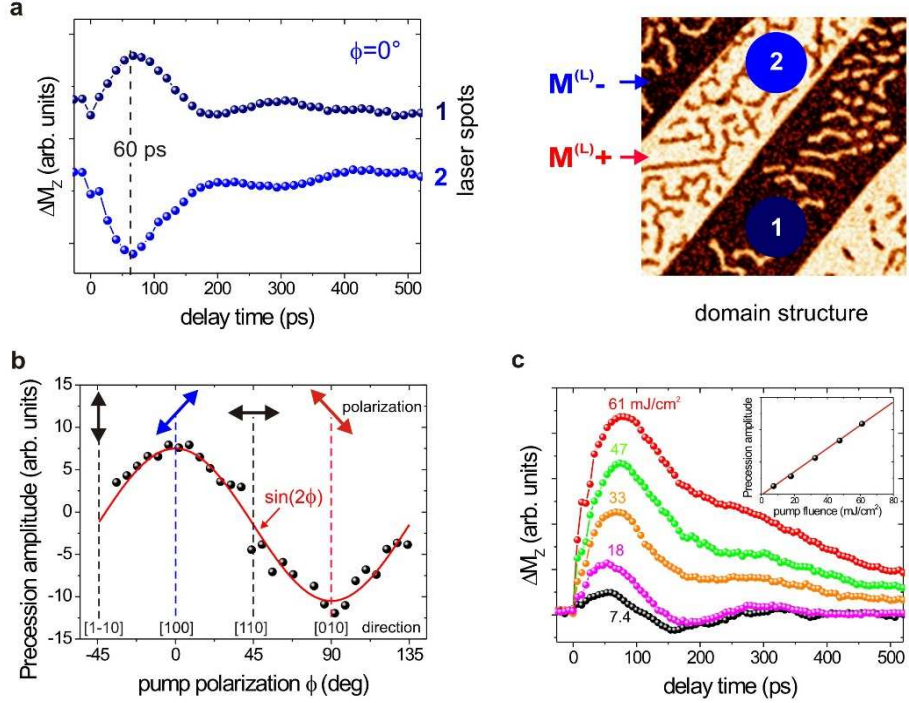

**Figure S1 | Time-resolved magnetization precession induced by the femtosecond pump pulses in YIG:Co film at room temperature.** The out-of-plane component of the magnetization $M_Z$ is detected with the help of time-resolved magneto-optical Faraday rotation. (**a**) (left) The laser-induced magnetization precession for the case when the light is polarized along the [100] orientation and the magnetization is either in $M^{(L)}+$ or in $M^{(L)}-$ state. (Right) the domain structure and the spots in which the dependences shown on the left panel were measured. (**b**) Dependence of the precession amplitude on the pump polarization $M^{(L)}+$ domain. Solid line is a fit by $sin(2\varphi)$ function. (**c**) The dynamics measured at different pump fluences $I$ in the range from 7.4 to 61 mJ cm$^{-2}$. The pump polarization was [100] direction. The inset shows the linear dependence of the precession amplitude on the pump fluence.

**Symmetry analysis of the all-optical switching under linear polarized light.** Let's choose the coordinate system in which the coordinate axes $x$, $y$ and $z$ are aligned along the crystallographic directions [100], [010] and [001], respectively. To explain the effect of light on the magnetic anisotropy, the energy of light-matter interaction should contain terms $\chi_{ijkl}E_iE_j^*M_kM_l$, where $\chi_{ijkl}$ is a polar fourth rank tensor, $E_i$ is the $i$-th component of the electric field of light, $E_j^*$ is the complex conjugate of the $j$-component of the electric field of light and $M_k$ is the $k$-th component of the magnetization[18]. In our experiment polarized light switches two domains with $M_y>0$ to two domains with $M_y<0$ and back. In all four states between which the light switches the magnetization in our experiment $M_x>0$. The terms for which $k=l$ do not depend on the sign of the magnetization components and thus cannot be responsible for the sign change of $M_y$. It is also seen that the switching does not depend on the sign of $M_z$, because the same pulse has opposite effects on $M_z$ in large and small domains. Hence if light is polarized along [100] or [010] direction, the part of the energy of the photo-induced magnetic anisotropy responsible for the switching can be written as

$$W_L(\mathbf{E},\mathbf{M})= \chi_{xxyx}E_xE_x^*M_yM_x + \chi_{xxxy}E_xE_x^*M_xM_y+ \chi_{yyxy}E_yE_y^*M_xM_y + \chi_{yyyx}E_yE_y^*M_yM_x \quad (1),$$

This energy must be considered as a term of thermodynamic potential minimization of which with respect to **M** allows to find the potential minimum and define the equilibrium orientation of the magnetization.

The expected point group for the (001) surface of the garnet is *4mm* for which[30] $\chi_{yyyx}=\chi_{xxxy}=\chi_{xxyx}=\chi_{yyxy}=0$. However, a non-zero magnetization along the [001] axis, i.e. $M_z$, effectively lowers the symmetry down to *4*. The symmetry of the real garnet films can be also lowered simply due to distortions during growth[25]. For the point group *4* $\chi_{yyyx}= -\chi_{xxxy}$ and $\chi_{xxyx}=-\chi_{yyxy}$ [30]. Assuming that $A=\chi_{xxxy}+\chi_{xxyx}$ Eq. (1) acquires an easier form

$$W_L(\mathbf{E},\mathbf{M})=A(E_xE_x^*M_xM_y - E_yE_y^*M_xM_y) \quad (2).$$

Assume that *A<0*. In this case, under illumination of the garnet with light linearly polarized along the [100] direction the thermodynamic potential will be minimized if $M_y>0$. Hence, such an excitation will promote switching of large white domains ($M^{(L)}+$) into large black ones ($M^{(L)}-$). Simultaneously, small black domains ($M^{(S)}-$) will be switched into small white ones ($M^{(S)}+$). If the light is polarized along the [010] axis, the thermodynamic potential is minimized when $M_y<0$ and it means that the photoexcitation will promote switching of the magnetization in the large and small domains back into the $M^{(L)}+$ and $M^{(S)}-$ states, respectively.

11